\documentclass[prd,reprint,showpacs,showkeys]{revtex4}
\usepackage{amsfonts}
\usepackage{amssymb}
\usepackage{amsmath}
\usepackage{graphicx}
\usepackage{float}
\usepackage[font={footnotesize,it}]{caption}

\begin{document}

\title{Absence of Buckling in Nerve Fiber}
\author{S. Habib Mazharimousavi}
\email{habib.mazhari@emu.edu.tr}
\author{M. Halilsoy}
\email{mustafa.halilsoy@emu.edu.tr}
\affiliation{Department of Physics, Faculty of Arts and Sciences, Eastern Mediterranean
University, Famagusta, North Cyprus via Mersin 10, Turkey}

\begin{abstract}
In this study we give a geometrical model which employs the smoothness of
nerve fibers as differentiable curves. We show that a nerve fiber may
encounter large curvature due to the possible helicial bending and hence it
could cause the fiber to buckle. However, its membrane structure provides a
mechanism, entirely geometrical to avoid it. To overcome the challenge of
emerging helix we project it into a plane.
\end{abstract}

\maketitle

\section{INTRODUCTION}

Biological applications of geometry, especially differential geometry, is an
attractive and newly expanding branch of science that explains common
grounds in both disciplines \cite{1, 2}. Differential geometry is known to
find vast application in geometrical formulation of Einstein's general
relativity. The metric structure, continuity, differentiability,
connections, curvature, torsion, etc. are well-known concepts that are
instrumental for the formulation of gravity as a geometrical theory. The
elements of geometry are lines, surfaces, volumes of any sort and their
product combinations to define the underlying topology. With the advent of
exotic objects such as black holes, wormholes, singularities, cosmic strings
and others which make subject matters for the present gravitational theory
it becomes natural to innovate similar concepts as different applications of
geometry elsewhere, such as biological systems. The geometry of cells which
is in essence a cylinder with circular or elliptical radial cross section
can be adapted to imitate axons (as fibers) and similar structures in a
biological system \cite{3}. It is these cylindrical structures through which
electric conduction occurs and the rules of diffusion between different
parts take place. The walls of these cells are membranes separating
different regions which are seriously effected by swellings and
non-isotropic deformations. Any such geometrical deformation manifests
significant changes in electric flow signal's voltages and in the underlying
inter cellular current distributions. It is natural therefore to associate
neurodegenerative diseases such as Alzheimer, Parkinson, HIV etc. \cite{4}
to such deformations.\newline
In a recent study \cite{3} a general cell's equation has been derived with
physical consequences whenever swellings took place in the circular cross-
sectional cylindrical models. The resulting diffusion equation contains
beside the geometrical term an external source term which was phrased as the
geometrical diffusion term. The cross section of cell, however, was
restricted only to the circular ones which didn't cover the
non-circular/elliptical shapes. Our principal aim in this study is to remove
this restriction and consider more general geometric cross- sections. In
other words, our cylindrical membranes will be a function of both the length
of the cell as well as the angular variable as a requirement to remove the
planar isotropy. The mathematical tool which we shall employ will be
analogous to Ref. \cite{3} where an appropriate Frenet-Serret triplet frame
will be constructed as our reference frame. The geometrical change of the
vector triplet, i.e. $\vec{T}=$ tangent vector, $\vec{N}=$ normal vector and
$\vec{B}=\vec{T}\times \vec{N}$, along the curve parametrized by arc length $%
s$ and axial angle $\phi $ will be satisfying the standard Frenet-Serret
equations \cite{5}. The axon's body in the nerve cell, for instance will be
expressed as a linear combination of the three vectors $\vec{T},\vec{N}$ and
$\vec{B}$. The surface of the membrane can be represented by

\begin{equation}
\vec{F}=\vec{\alpha}(s)+\nu (s,\phi )\vec{N}+\beta (s,\phi )\vec{B},
\end{equation}%
where $\vec{\alpha}(s)$ (along $\vec{T}$), $\nu (s,\phi )$ and $\beta
(s,\phi )$ are functions of their arguments, which are to be determined from
the geometrical ansatz model. From the first fundamental form $g_{ij}$ of
the surface we derive the second fundamental form known also as the
extrinsic curvature $K_{ij}.$ The latter describes how a given surface is
embedded in the $3-$dimensional Euclidean space in which we can embed our
surface. The determinant and trace of the tensor $K_{ij}$ are useful
geometrical quantities that guide us in analyzing the geometrical structure
of membranes. Beside these, the cross- sectional area of a cylindrical
structure is a useful parameter in studying buckle formation. From the
principles of differential geometry it is known that at a corner point where
two lines (or surfaces) intersect at an angle the vectors are not
differentiable, which amounts to a singular point/line. Namely, approach of
tangent vectors from different directions do not match in slopes. We shall
borrow the same terminology in application to the fibers to eliminate such
singularities which will amount ultimately to the absence of buckling in
structures such as fibers. Let us recall that the concept of singularity in
physics is an infamous one and in a biological system specifically is
totally non-acceptable. Geometrically we employ the bending energy of the
fluid membrane \cite{2} which is expressed in Gaussian and extrinsic
curvatures. That must be finely-tuned in order to have a tractable model.
The extremals of such an energy yield the well-known helix, for instance,
wherever planar isometry is imposed. For a more general treatment, however,
we admit that isotropy in this cross-section of the fiber structure must be
relaxed. To this end we appeal to the Hamiltonian formalism developed by
Helfrich and search for the extremal (the minimum) energy conditions.
Tubular fibers with circular cross-section is relatively a simple problem,
however, extending this to non-circular cross-sections seems to be
challenging. Although this makes the ultimate aim of the present article we
admit that we were able to solve the problem under restricted conditions.
Organization of the paper goes as follows. In section II we introduce our
mathematical formalism and study the case of circular cross-sections.
Non-circular projection of a helicial bending is analyzed in Section III.
The paper is completed with our Conclusion in Section IV.

\section{THE FORMALISM}

We start with the mathematical model of an axon embedded in a three
dimensional Euclidean space given by the recent work of Lopez-Sanchez and
Romero \cite{3}. In this model an axon is built on a three dimensional
continuous curve $\overrightarrow{\mathbf{\alpha }}\left( s\right) $ in
which $s$ is the arc length parameter of the curve. Using the above
parametrization, one uses the Frenet-Serret coordinate system with the
following three unit vectors%
\begin{equation}
\mathbf{T}=\frac{d\mathbf{\alpha }\left( s\right) }{ds},\mathbf{N}=\frac{d%
\mathbf{T}/ds}{\left\Vert d\mathbf{T}/ds\right\Vert },\mathbf{B=T\times N},
\end{equation}%
in which $\mathbf{T}$ is the unit tangent vector to the curve, $\mathbf{N}$
is called unit normal vector, $\mathbf{B}$ is known as the unit binormal
vector and $\left\Vert d\mathbf{T}/ds\right\Vert $ is the magnitude of $d%
\mathbf{T}/ds$. The Frenet-Serret unit normals satisfy%
\begin{equation}
\frac{d}{ds}\left(
\begin{array}{r}
\mathbf{T} \\
\mathbf{N} \\
\mathbf{B}%
\end{array}%
\right) =\left(
\begin{array}{rrr}
0 & \kappa \left( s\right)  & 0 \\
-\kappa \left( s\right)  & 0 & \tau \left( s\right)  \\
0 & -\tau \left( s\right)  & 0%
\end{array}%
\right) \left(
\begin{array}{r}
\mathbf{T} \\
\mathbf{N} \\
\mathbf{B}%
\end{array}%
\right),
\end{equation}%
in which $\kappa \left( s\right) $ and $\tau \left( s\right) $ are the
curve's curvature and torsion, respectively. Following Ref. \cite{3} we
construct the surface of the axon, using $\overrightarrow{\mathbf{\alpha }}%
\left( s\right) $ as the skeleton of the model and two new functions $\nu
\left( s,\phi \right) $ and $\beta \left( s,\phi \right) ,$ which give the
normal extensions to the skeleton of the axon's body expressed by%
\begin{equation}
\overrightarrow{\mathcal{F}}:=\overrightarrow{\mathbf{\alpha }}\left(
s\right) +\nu \left( s,\phi \right) \mathbf{N+}\beta \left( s,\phi \right)
\mathbf{B.}
\end{equation}%
Herein both $\nu \left( s,\phi \right) $ and $\beta \left( s,\phi \right) $
are analytic in terms of $s$ and $\phi $ whereas $\phi $ itself is an angle
measured from a given reference point. Having the surface of the axon
defined and parametrized as in Eq. (3) one adopts the first fundamental form
given by%
\begin{equation}
g_{ij}=\left(
\begin{array}{rr}
E & F \\
F & G%
\end{array}%
\right),
\end{equation}%
where $E=\frac{\partial \overrightarrow{\mathcal{F}}}{\partial s}.\frac{%
\partial \overrightarrow{\mathcal{F}}}{\partial s},G=\frac{\partial
\overrightarrow{\mathcal{F}}}{\partial \phi }.\frac{\partial \overrightarrow{%
\mathcal{F}}}{\partial \phi }$ and $F=\frac{\partial \overrightarrow{%
\mathcal{F}}}{\partial s}.\frac{\partial \overrightarrow{\mathcal{F}}}{%
\partial \phi }$ are given by%
\begin{equation}
E=\left( 1-\kappa \left( s\right) \nu \right) ^{2}+\left( \nu _{,s}-\tau
\left( s\right) \beta \right) ^{2}+\left( \beta _{,s}+\tau \left( s\right)
\nu \right) ^{2},
\end{equation}%
\begin{equation}
F=\left( \nu _{,s}-\tau \left( s\right) \beta \right) \nu _{,\phi }+\left(
\beta _{,s}+\tau \left( s\right) \nu \right) \beta _{,\phi }
\end{equation}%
and%
\begin{equation}
G=\left( \nu _{,\phi }\right) ^{2}+\left( \beta _{,\phi }\right) ^{2},
\end{equation}%
in which $\left( .\right) _{,x}=\frac{\partial \left( .\right) }{\partial x}.
$ For the parametrized surface $\overrightarrow{\mathcal{F}}$ one defines
\begin{equation}
\hat{n}=\frac{\frac{\partial \overrightarrow{\mathcal{F}}}{\partial s}\times
\frac{\partial \overrightarrow{\mathcal{F}}}{\partial \phi }}{\left\Vert
\frac{\partial \overrightarrow{\mathcal{F}}}{\partial s}\times \frac{%
\partial \overrightarrow{\mathcal{F}}}{\partial \phi }\right\Vert }
\end{equation}%
to be the unit normal vector at any point of the surface explicitly given by%
\begin{equation}
\hat{n}=\frac{1}{g}\left[ \left( \nu _{,s}\beta _{,\phi }-\beta _{,s}\nu
_{,\phi }-\frac{1}{2}\tau \left( \nu ^{2}+\beta ^{2}\right) _{,\phi }\right)
\mathbf{T}-\left( 1-\kappa \nu \right) \beta _{,\phi }\mathbf{N}+\left(
1-\kappa \nu \right) \nu _{,\phi }\mathbf{B}\right],
\end{equation}%
in which $g=\det \left( g_{ij}\right) =EG-F^{2}.$ The element of area on the
surface is given by%
\begin{equation}
dA=dsd\phi \sqrt{g}=dsd\phi \sqrt{\left( \nu _{,s}\beta _{,\phi }-\beta
_{,s}\nu _{,\phi }-\frac{1}{2}\tau \left( \nu ^{2}+\beta ^{2}\right) _{,\phi
}\right) ^{2}+\left( 1-\kappa \nu \right) ^{2}\left( \beta _{,\phi }^{2}+\nu
_{,\phi }^{2}\right) } . \nonumber
\end{equation}%
The second fundamental form i.e., the extrinsic curvature tensor is defined
by%
\begin{equation}
K_{ij}=\frac{\partial ^{2}\overrightarrow{\mathcal{F}}}{\partial
u^{i}\partial u^{j}}.\hat{n},
\end{equation}%
in which $u^{i}\in \left\{ s,\phi \right\} .$ The general expression of the
curvature tensor's components are too long to be given here. Our ultimate
reason is to calculate the extrinsic curvature tensor and to investigate the
behavior of the two scalar invariants of the surface, namely the Gaussian
curvature and the total curvature which are given respectively by $K=\det
K_{i}^{j}$ and $H=K_{i}^{i}=TrK_{i}^{j}.$ We define the smooth surface to
have analytic / differentiable $K$ and $H.$

To make our analysis practical we consider some specific, yet important
cases. The first case which has also been considered in Ref. \cite{3} is the
circular cross sectional axons defined by $\nu \left( s,\phi \right)
=R\left( s\right) \cos \phi $ and $\beta \left( s,\phi \right) =R\left(
s\right) \sin \phi $ in which $R\left( s\right) $ refers to the radius of
the circular cross section.

\subsection{Circular Cross Section}

The Gaussian and total curvature for this circular cross sectional axon are
found to be%
\begin{equation}
K=\frac{2\left( -\omega ^{2}R^{\prime \prime }+\left( \left( 3\omega
+1\right) \cos \phi -\kappa R\right) \kappa R^{\prime 2}+R\omega \left(
\kappa ^{\prime }\cos \phi +\kappa \tau \sin \phi \right) R^{\prime }+\kappa
\omega ^{3}\cos \phi \right) }{R\left( R^{\prime 2}+\omega ^{2}\right) ^{2}}
\end{equation}%
and%
\begin{equation}
H=\frac{\omega RR^{\prime \prime }-\left( 2+3\omega \right) R^{\prime
2}-R^{2}\left( \kappa ^{\prime }\cos \phi +\tau \kappa \sin \phi \right)
R^{\prime }-\omega ^{2}\left( 2\omega +1\right) }{R\left( R^{\prime
2}+\omega ^{2}\right) ^{3/2}},
\end{equation}%
in which $\omega =R\kappa \cos \phi -1$ and a prime stands for the
derivative with respect to $s.$ Let's add that Eq. (11) is the generalized
expression of Eq. (15) of Ref. \cite{3} where $R\left( s\right)
=R_{0}=const. $.

As mentioned above the smooth surface implies that the Gaussian and total
curvature are finite everywhere on the surface. Hence, Eq. (11) and Eq. (12)
manifestly imply that to have the surface differentiable / analytic
everywhere a nonzero $R^{\prime }\left( s\right) $ is a sufficient condition
but not necessary indeed. In other words, if $R^{\prime }\left( s\right) $
is nonzero at a point then both $H$ and $K$ are finite but if $R^{\prime
}\left( s\right) $ vanishes at a point then $\omega $ must be nonzero at
that point. This, however, imposes that $R\kappa <1.$ We note that, imposing
the strong constraint $R\kappa <1$ is also sufficient to have the manifold
to be differentiable, but this is not necessary. Section B considers the
particular case as follows.

\subsection{$R\left( s\right) =R_{0}=const.$}

Now, let's assume that $R\left( s\right) =R_{0}$ and $R_{0}\kappa \geq 1$
where one finds
\begin{equation}
K=-\frac{\kappa \left( s\right) \cos \phi }{R_{0}\omega }
\end{equation}%
and
\begin{equation}
H=\frac{\left( 2\omega +1\right) }{R_{0}\omega },
\end{equation}%
which are not analytic at the critical $\phi $ where $\omega =R\kappa \cos
\phi -1=0,$ and consequently the axon buckles at those points. Based on the
fact that the Gaussian curvature of the surface of the axon is a continuous
function of $\phi $ (provided $R_{0}\kappa \left( s\right) <1$), it admits
two local extrema at $\phi =0$ and $\phi =\pi $ such that
\begin{equation}
\left. K\right\vert _{\phi =0}=-\frac{\kappa \left( s\right) }{R_{0}\left(
1-R_{0}\kappa \left( s\right) \right) }
\end{equation}%
and
\begin{equation}
\left. K\right\vert _{\phi =\pi }=\frac{\kappa \left( s\right) }{R_{0}\left(
1+R_{0}\kappa \left( s\right) \right) }.
\end{equation}%
Furthermore, we expect the two extremals to occur at points $\phi =0$ and $%
\phi =\pi $. Hence, one finds out that the reference of $\phi $ is the point
$\phi =0$ as the innermost point toward the center of curvature. When we
relax the constraint on $R_{0}\kappa \left( s\right) $ we observe that for
the case $R_{0}\kappa \left( s\right) =1$ there would be only a single point
i.e., $\phi =0$ where the axon buckles but for $R_{0}\kappa \left( s\right)
>1$ there are two points where the Gaussian curvature diverges which are
given by%
\begin{equation}
\phi =\cos ^{-1}\frac{1}{R_{0}\kappa \left( s\right) }=\pm \phi _{b},
\end{equation}%
in which $0<\phi _{b}<\frac{\pi }{2}$ and in the limit where $R_{0}\kappa
\left( s\right) $ gets very large the value of the angle $\phi _{b}$
approaches to $\frac{\pi }{2}.$ We note that the physical interpretation of $%
\kappa \left( s\right) \geq \frac{1}{R_{0}}$ is that the center of curvature
of the curve $\mathbf{\alpha }\left( s\right) $ is inside the body of the
axon.

If the surface of the axon were not made of the lipid membrane, the actual
situation would be as we described above. But the fact is that the lipid
membrane of the axon leaves its radius to be a function of $s$ in the
vicinity of the large curvature of the curve $\mathbf{\alpha }\left(
s\right) $ (i.e., the place where $\kappa \left( s\right) $ is very large).
Hence from Eq. (11) one finds that $K$ remains finite and this results in a
smooth transition without a buckling. It is remarkable to observe that the
term which rescues the surface from buckling is not $R$ itself but its first
derivative i.e., $R^{\prime }.$ Therefore there would not be any restriction
on the radius of the axon in general, no matter what would be their passage
condition within the body. In the following section we consider the case of
an axon with non-circular cross section which generalizes the analysis of
Ref. \cite{3}.

\section{CURVE OF BENDING}

In this section we assume that the axon's radius is constant i.e., $R\left(
s\right) =R_{0}$ and the axon's curvature satisfies the condition $\kappa
\left( s\right) R_{0}<1.$ Upon these assumptions we look for the equation of
the curve which joins two straight line parts of the axon.

The total energy of a lipid bilayer is expressed by the Helfrich Hamiltonian
(see Ref. \cite{6})%
\begin{equation}
E=\int dA\left\{ \sigma +\frac{1}{2}k_{1}\left( H-H_{0}\right)
^{2}+k_{2}K\right\} ,
\end{equation}%
where $\sigma $ is the surface tension, $H_{0}$, $H$ and $K$ are
spontaneous, total and Gaussian curvatures, respectively, $k_{1}$ and $k_{2}$
are bending and the Gaussian curvature moduli. The integral is taken over
the whole surface of bilayer membrane. Therefore, as can be observed from
this Hamiltonian, the energy of a lipid membrane is a function of bilayer's
geometrical / topological properties, emerged in shape of its curvatures.
Here $\sigma $ is assumed to be an independent thermodynamic surface
tension, which reflects the chemical potential of lipids, and therefore is
taken as a constant. Also, it is worth mentioning that except for some
certain types of lipid membranes \cite{7,8} in case the two sides of a
bilayer are not distinguishable, the spontaneous curvature $H_{0}$ vanishes
\cite{9}. In the following sections, when the variation of Helfrich
Hamiltonian is evaluated, $\sigma $ is treated as a constant and the sides
of the membrane are assumed indistinguishable. Upon these assumptions and
our results in previous section i.e.,
\begin{equation}
K=\frac{\kappa \left( s\right) \cos \phi }{R_{0}\left( R_{0}\kappa \left(
s\right) \cos \phi -1\right) }
\end{equation}%
and
\begin{equation}
H_{T}=\frac{1-2R_{0}\kappa \left( s\right) \cos \phi }{R_{0}\left(
1-R_{0}\kappa \left( s\right) \cos \phi \right) },
\end{equation}%
with the areal element%
\begin{equation}
dA=dsd\phi \sqrt{g}=dsd\phi R_{0}\left( 1-R_{0}\kappa \left( s\right) \cos
\phi \right),
\end{equation}%
one finds the bending energy of the axon given by
\begin{equation}
E=\int_{0}^{s}ds^{\prime }\int_{0}^{2\pi }\frac{-\xi _{1}R_{0}^{2}\kappa
\left( s^{\prime }\right) ^{2}\cos ^{2}\phi +2\xi _{2}\kappa \left(
s^{\prime }\right) \cos \phi +\xi _{3}}{2R_{0}\left( R_{0}\kappa \left(
s^{\prime }\right) \cos \phi -1\right) }d\phi ,
\end{equation}%
in which
\begin{equation}
\xi _{1}=R_{0}^{2}\left( 2\sigma +k_{1}H_{0}^{2}\right) +2R_{0}\left(
k_{2}-2k_{1}H_{0}\right) +4k_{1},
\end{equation}%
\begin{equation}
\xi _{2}=R_{0}^{2}\left( 2\sigma +k_{1}H_{0}^{2}\right) +R_{0}\left(
k_{2}-3k_{1}H_{0}\right) +2k_{1}
\end{equation}%
and%
\begin{equation}
\xi _{3}=-R_{0}^{2}\left( 2\sigma +k_{1}H_{0}^{2}\right)
+2k_{1}R_{0}H_{0}-2k_{1}.
\end{equation}

The integral on $\phi $ can be calculated by using the Residual theorem from
complex analysis. Let's introduce%
\begin{equation}
I=\int_{0}^{2\pi }\frac{-\xi _{1}R_{0}^{2}\kappa \left( s^{\prime }\right)
^{2}\cos ^{2}\phi +2\xi _{2}\kappa \left( s^{\prime }\right) \cos \phi +\xi
_{3}}{2R_{0}\left( R_{0}\kappa \left( s^{\prime }\right) \cos \phi -1\right)
}d\phi
\end{equation}%
such that a change of variable of the form $z=e^{i\phi }$ yields%
\begin{equation}
I=\frac{i}{4}\oint\limits_{C}\frac{\xi _{1}R_{0}^{2}\kappa \left( s^{\prime
}\right) ^{2}\left( 1+z^{2}\right) ^{2}-4\xi _{2}R_{0}\kappa \left(
s^{\prime }\right) z\left( 1+z^{2}\right) -4\xi _{3}z^{2}}{R_{0}z^{2}\left(
R_{0}\kappa \left( s^{\prime }\right) \left( 1+z^{2}\right) -2z\right) }d\phi,
\end{equation}%
in which the contour $C$ is the unit circle. This integral is equal to
\begin{equation}
I=2i\left( 2\pi i\sum a_{-1}\right),
\end{equation}%
where $a_{-1}$ are the residue of the poles inside contour $C.$ There are
three poles located at $z_{01}=0,$ $z_{02}=\frac{1-\sqrt{1-e^{2}}}{e}$ and $%
z_{03}=\frac{1+\sqrt{1-e^{2}}}{e}$ which upon the choice $e=R_{0}\kappa
\left( s^{\prime }\right) <1,$ only $z_{01},z_{02}$ of order $m=2$ and $m=1,$
respectively, are located inside the contour with residues
\begin{equation}
a_{-1}\left( z_{0}=0\right) =\frac{i\left( \xi _{1}-2\xi _{2}\right) }{2R_{0}%
}
\end{equation}%
and
\begin{equation}
a_{-1}\left( z_{0}=\frac{1-\sqrt{1-e^{2}}}{e}\right) =-\frac{i\left( \xi
_{1}-2\xi _{2}-\xi _{3}\right) }{2R_{0}\sqrt{1-e^{2}}},
\end{equation}%
respectively. Finally we find%
\begin{equation}
I=\frac{2\pi }{\sqrt{1-e^{2}}}
\end{equation}%
and consequently the energy integral reduces to%
\begin{equation}
E=\pi \int_{0}^{s}ds^{\prime }\left\{ \frac{k_{1}}{R_{0}\sqrt{%
1-R_{0}^{2}\kappa \left( s^{\prime }\right) ^{2}}}+\left( R_{0}\left(
2\sigma +k_{1}H_{0}^{2}\right) -2k_{1}H_{0}\right) \right\} .
\end{equation}%
$E$ is a functional depending on the function $\kappa \left( s^{\prime
}\right) $ only such that we are looking for a specific $\kappa \left(
s^{\prime }\right) $ which makes the functional $E$ stationary. Using the
calculus of variation or the Euler equation we find
\begin{equation}
\kappa \left( s^{\prime }\right) =\kappa _{0},
\end{equation}%
in which $\kappa _{0}$ is a constant. Hence, the curve of the bending is a
constant-curvature curve. Here the boundary conditions are imposed by the
two curves joined to this constant-curvature at the initial and final points
of the curve.

Let's assume that the curve is planar and joins two straight parts of the
axon located on the same plane as projection of the curve itself. In this
case one finds the total energy to be%
\begin{equation}
E_{Plane}=\pi s\left\{ \frac{k_{1}}{R_{0}\sqrt{1-R_{0}^{2}\kappa _{0}^{2}}}%
+\left( R_{0}\left( 2\sigma +k_{1}H_{0}^{2}\right) -2k_{1}H_{0}\right)
\right\},
\end{equation}%
in which $s$ is the arc length of the circle of radius $r=\frac{1}{\kappa
_{0}}.$ Hence, $s=\psi r,$ where $\psi $ is the angle of total bending and
is a constant dictated by the initial conditions. As a result we find the
total energy of the curve to be%
\begin{equation}
E_{Plane}=\pi \psi r\left\{ \frac{k_{1}}{R_{0}\sqrt{1-\frac{R_{0}^{2}}{r^{2}}%
}}+\left( R_{0}\left( 2\sigma +k_{1}H_{0}^{2}\right) -2k_{1}H_{0}\right)
\right\},
\end{equation}%
where we have used $\kappa _{0}=\frac{1}{r}.$ The final step is to find $r$
such that $E_{Plane}$ is stationary. Here $E_{Plane}$ is a function of $r$
only and upon $\frac{dE_{Plane}}{dr}=0$ one finds the radius of the bending.
Two particular cases are considered in the sequel.

\subsection{Case 1}

Let's assume also that
\begin{equation}
R_{0}\left( 2\sigma +k_{1}H_{0}^{2}\right) -2k_{1}H_{0}=0
\end{equation}%
which yields%
\begin{equation}
E_{Plane}=\frac{k_{1}\pi \psi r}{R_{0}\sqrt{1-\frac{R_{0}^{2}}{r^{2}}}}
\end{equation}%
whose minimum occures at
\begin{equation}
r=\sqrt{2}R_{0}.
\end{equation}%
Therefore, we have found the radius of the circle under which the axons bend
such that the bending energy of the curling becomes an extremum which can
easily be shown to be a minimum.

\subsection{Case 2}

For a more general case where
\begin{equation}
R_{0}\left( 2\sigma +k_{1}H_{0}^{2}\right) -2k_{1}H_{0}=\lambda \neq 0
\end{equation}%
the stationary radius is found to be%
\begin{equation}
r=R_{0}\sqrt{\frac{\sqrt[3]{4k_{1}^{2}\alpha }}{6\left( R_{0}^{2}\lambda
^{2}-k_{1}^{2}\right) }-\frac{\left( 3R_{0}^{2}\lambda
^{2}-4k_{1}^{2}\right) k_{1}^{2}\sqrt[3]{16}}{6\left( R_{0}^{2}\lambda
^{2}-k_{1}^{2}\right) \sqrt[3]{4k_{1}^{2}\alpha }}+\frac{3R_{0}^{2}\lambda
^{2}-4k_{1}^{2}}{3\left( R_{0}^{2}\lambda ^{2}-k_{1}^{2}\right) }},
\end{equation}%
in which
\begin{equation}
\alpha =\left( 3\lambda ^{3}R_{0}^{3}-3\lambda R_{0}k_{1}^{2}\right) \sqrt{3}%
\sqrt{27R_{0}^{2}\lambda ^{2}-32k_{1}^{2}}+27R_{0}^{4}\lambda
^{4}-45R_{0}^{2}\lambda ^{2}k_{1}^{2}+16k_{1}^{4}.
\end{equation}%
Since we assumed a priori that $\lambda \neq 0,$ we don't take the limit $%
\lambda \rightarrow 0$ in this particular solution.

\section{CONCLUSION}

In a rough analogy nerve fibers are tubes with cylindrical topology. Since
physics aims to describe every kind of systems, including biological ones,
in mathematical terms, differential geometry becomes the right tool for this
purpose. In order to have buckling in the fibers these must be singularities
in the underlying geometrical structure. Divergence in the curvature scalars
is not the only criterion that determines such singularities in the present
problem. Instead, differentiability analysis for each curve on the tubular
manifold determines the regularity and therefore absence of buckling. Recall
that differentiability implies continuity but the converse statement need
not be true. Derivative of radius function in terms of arclength / angle
exists everywhere, which can be interpreted as an indication of regularity,
or absence of buckling. We aim fibers with non-circular cross sections and
we show, with reference to the minimal energy of the Helfrich Hamiltonian,
that fibers don't buckle. In doing this we project the helix structure into
the plane and investigate the continuity of the tangent vectors everywhere.
In particular examples we attain results, leaving more general cases to
future studies in which possible double-helix structures may also be taken
into account.

\end{document}